\documentclass[12pt]{article}
\title{ Nonperturbative quark dynamics in  a baryon}
\author{Yu.A.Simonov\\
 Jefferson
Laboratory, Newport News, VA 23606, USA, and\\
 State Research
Center\\Institute of Theoretical and Experimental Physics, \\
Moscow, 117218 Russia}
 \date{}
\newcommand{\beq}{\begin{eqnarray}}
 \newcommand{\eeq}{\end{eqnarray}}
\newcommand{\be}{\begin{equation}}
 \newcommand{\ee}{\end{equation}}

 \def\la{\mathrel{\mathpalette\fun <}}

\def\fun#1#2{\lower3.6pt\vbox{\baselineskip0pt\lineskip.9pt
\ialign{$\mathsurround=0pt#1\hfil ##\hfil$\crcr#2\crcr\sim\crcr}}}

\newcommand{{\SD}}{\rm SD}
\newcommand{\pp}{\prime\prime}

\newcommand{\ver}{\mbox{\boldmath${\rm r}$}}
\newcommand{\vesig}{\mbox{\boldmath${\rm \sigma}$}}
\newcommand{\vedelta}{\mbox{\boldmath${\rm \delta}$}}
\newcommand{\veP}{\mbox{\boldmath${\rm P}$}}
\newcommand{\vep}{\mbox{\boldmath${\rm p}$}}
\newcommand{\veq}{\mbox{\boldmath${\rm q}$}}
\newcommand{\vez}{\mbox{\boldmath${\rm z}$}}

\newcommand{\veL}{\mbox{\boldmath${\rm L}$}}
\newcommand{\veR}{\mbox{\boldmath${\rm R}$}}

\newcommand{\ven}{\mbox{\boldmath${\rm n}$}}
\newcommand{\veu}{\mbox{\boldmath${\rm u}$}}
\newcommand{\vev}{\mbox{\boldmath${\rm v}$}}

\newcommand{\vexi}{\mbox{\boldmath${\rm \xi}$}}
\newcommand{\veta}{\mbox{\boldmath${\rm \eta}$}}
\newcommand{\veB}{\mbox{\boldmath${\rm B}$}}

\newcommand{\veE}{\mbox{\boldmath${\rm E}$}}

\newcommand{\llan}{\langle\langle}
\newcommand{\rran}{\rangle\rangle}
\newcommand{\lan}{\langle}
\newcommand{\ran}{\rangle}
\begin{document}
\maketitle

\begin{abstract}

Field-correlator method is used to calculate nonperturbative
dynamics of quarks in a baryon. General expression for the 3q
Green's function is obtained using Fock-Feynman-Schwinger
(world-line) path integral formalism, where all dynamics is
contained in the 3q Wilson loop with spin-field insertions. Using
lowest cumulant contribution for  Wilson loop one obtains an
$Y$-shaped string with a deep hole at the string junction
position. Using einbein formalism for the quark kinetic terms one
automatically obtains constituent quark masses, calculable
through the string tension. Resulting effective action for 3q plus
$Y$-shaped strings is quantized in the path-integral formalism to
produce two versions of Hamiltonian,one -in the c.m. and another in the
light-cone system.
Hyperspherical formalism is used to calculate masses and wave
functions.Simple estimates in lowest approximation  yield baryon
masses in good agreement with experiment without fitting
parameters.
 \end{abstract}

 \section{Introduction}

Baryons are for a long time an object of an intensive theoretical
study \cite{1}-\cite{11}. Both the perturbative dynamics  and
confinement interaction were considered decades ago
\cite{1}-\cite{5} and a series of  papers of Isgur and
collaborators \cite{7,8} has  enlightened the structure of the
baryon spectrum in good general agreement with experiment. In
those works dynamics was considered  as a QCD motivated and
relativistic effects in kinematics have been accounted for.
Recently a more phenomenological approach based on large $N_c$
expansion for baryons \cite{12}-\cite{15} was applied to baryon
spectra \cite{16,17} and clearly demonstrated the most important
operators forming the spectrum of 70- plet.

Summing up the  information from the quark-based model one has a
picture of baryon spectra with basically oscillator-type spectrum,
modified by presence of spin-dependent forces and other
corrections. E.g. the orbital excitation with $\Delta L=1$ "costs"
around 0.5 GeV, while the radial one (actually two types) amounts
to around 0.8 GeV. Moreover, hyperfine splitting, which in
experiment is large (for $\Delta-N $ system it is 0.3 GeV) is
underestimated using perturbative forces with $\alpha_s\approx 0.4$ and
spin-orbit splitting typically small in experiment, needs some
special cancellations in theory \cite{18}.

Moreover, some states cannot be well explained in the standard
quark models. A good example is $N^*(1440)$, which is too low to
be simply a radial excitation and moreover its experimental
electroproduction amplitudes \cite{19} are in evident conflict
with theory \cite{20}.

This example is probably not unique, and one can notice an
interesting pattern in "radial" excitations of $N,\Delta, \Lambda$
and $\sum$: in all cases three lowest states $M_1, M_2,M_3$ have
intervals $\Delta_1\equiv M_2-M_1\sim 400\div 500$ MeV,
$\Delta_2\equiv M_3-M_1\sim 600-700$ MeV.

Namely, in \begin{center}
 \vspace{1cm}
\begin{tabular}{|c|c|c|c|}
$N(939)$& $\Delta(1232)$&$\Lambda(1116)$&$ \sum(1193)$\\
$N(1440)$& $\Delta(1600)$&$\Lambda(1600)$&$ \sum(1660)$\\
$N(1710)$& $\Delta(1920)$&$\Lambda(1810)$&$ \sum(1770)$\\ &&&$
\sum(1880)$\\\hline $\Delta_1=500,$&$
\Delta_1=370,$&$\Delta_1=480, $ &$\Delta_1=470,$\\
$\Delta_2=770;$&$\Delta_2=690;$&$\Delta_2=700;
$&$\Delta_2=600$\end{tabular} \vspace{1cm} \end{center}
 one can
see that $\Delta_2$ corresponds to usual radia
l excitation, while
the energy interval $\Delta_1$ cannot be explained in a
simple way in standard quark models.

One can say more about  difficulties with the interpretation of
the $\Lambda(1405)$ state, quantitative descriptions of $\Delta N$
transitions etc. \cite{19,20}.

In this  situation it sounds reasonable to apply new dynamical
approaches, which are directly connected to the basic QCD
Lagrangian and where all approximations can be checked both
theoretically and numerically on the lattice.

Here belongs the field correlator method (FCM) started in
\cite{21,22} (for a review see \cite{23} and for more dynamical
applications  \cite{24}). It is aimed at expressing all
observables in terms of gauge-invariant field correlators. Its use
is largely facilitated by recent observation on the  lattice
\cite{25, 26}, that the lowest bilocal correlator gives dominant
contribution to the quark-antiquark forces, while higher
correlators contribute around 1\%. The use of FCM for meson
spectra \cite{27,28} has shown, that gross features of spectra can
be calculated through only string tension, while fine and
hyperfine structure require the knowledge of another
characteristics of bilocal correlator --
 the gluon correlation length \cite{29} which is known from
 lattice data \cite{30} and analytic calculations \cite{31,32}.

 In dynamical applications of the FCM to baryon spectra two
 different schemes are used presently; the relativistic
 Hamiltonian method (RHM) and the method of Dirac orbitals.
 The first was suggested  in \cite{33,34} and used  for baryon
 Regge trajectories in \cite{35,36} and for magnetic moments in
 \cite{37}. The second method was suggested in \cite{38} and
 exploited to calculate baryon magnetic moments in \cite{39}.

Recently an important element was added to the RHM for baryons,
namely all spin-dependent forces between quarks have been
calculated in the same Gaussian approximations \cite{40}. To
finalize the RHM for baryons one still needs to construct the full
baryonic Hamiltonian taking into account the energy of string
motion, nonperturbative self-energy corrections \cite{41} etc. The
present paper is aimed at the fulfilling this task. It contains the
detailed derivation of the full baryonic Hamiltonian both in the
c.m. and  in the light-cone system of coordinates, simple
estimates of spectra for spin-averaged masses and a preparatory
discussion of future explicit detailed calculations.

The paper is organized  as follows. In section 2 the $3q$ Green's
function is written down using the Fock-Feynman-Schwinger
representation \cite{42},in section 3 the c.m. and relative
coordinates are introduced and the einbein field $\mu(t)$ is
introduced which will give rise to the quark constituent mass. In
section 4 the resulting effective action is quantized and the full
c.m. Hamiltonian is explicitly written down.

Section 5 is devoted to the derivation of the light-cone
Hamiltonian, its physical interpretation and correspondence to the
partonic model.

In section 6 the construction of the  baryon wave function is
discussed, expansion of its coordinate part into a sum of
hyperspherical harmonics and analytic estimates of spin-averaged
spectra.

Spin-dependent forces are discussed in section 7 and section 8 is
devoted to conclusions.

\section{Baryon Green's function}

One can define initial and final states of  baryon as a
superposition of 3q states

\be
\Psi_{in,out} (x^{(1)}, x^{(2)}, x^{(3)}) = \Gamma_{in,out}
e_{a_1a_2a_3}\Pi \psi^{a_i}_{\gamma_i}(x^{(i)}, x^{(0)})\label{1}
\ee
 where $a_i$ are color indices, while
$\gamma_i$ contain both flavour and Dirac indices, and  a sum over
appropriate combinations of these last indices is assumed with
$\Gamma$ as coefficients. Moreover $x^{(i)}, x^{(0)}$ are coordinates
of quarks and of the string junction respectively.

The 3q Green's function can be written as
\be
G_{3q}(x^{(i)}|y^{(k)}) =\lan tr_Y \Gamma_{out}\prod^3_{i=1}
S^{(i)}_{a_i b_i} (x^{(i)}, y^{(i)}) \Gamma_{in}\ran\label{2}
 \ee
where we have neglected the quark determinant and defined
\be
tr_Y=\frac16 \sum_{a_i, b_i} e_{a_1a_2 a_3} e_{b_1b_2b_3}\label{3}
\ee
 and $S^{(i)}(x^{(i)}, y^{(i)})$ is the quark Green's function
in the external gluonic fields (vacuum and perturbative gluon
exchanges). For the latter one can use the exact
Fock-Feynman-Schwinger (FFS) form \cite{22,24,42}
\be
S(x,y) =(m-\hat D) \int^\infty_0 ds (Dz)_{xy} e^{-K} W_z(x,y) \exp
g\int^s_0 \sigma_{\mu\nu} F_{\mu\nu} (z(\tau)) d\tau\label{4} \ee
where $W_z$ is the phase factor along the contour
 $C_z(x,y)$ starting at $y$ and finishing at $x$,
 which goes along trajectory which is integrated in
$(Dz)_{xy}, K=m^2s+\frac14 \int^s_0 \dot{z}^2 d\tau$
 \be
 W_z(x,y) = P\exp ig \int^x_y A_\mu dz_\mu\label{5}
 \ee
 and $P$ is the ordering operator,
 while
 \be
 \sigma_{\mu\nu} F_{\mu\nu} = \left ( \begin{array} {ll}
 \vesig\veB,&\vesig \veE\\
  \vesig\veE,&\vesig \veB
  \end{array}\right )
  \label{6}
  \ee

Here $ \veE, \veB $ are color-electric and color-magnetic fields respectively.
The average over gluon fields, implied in (\ref{2})
 by angular brackets, is  convenient to perform
 after the  nonabelian Stokes theorem is applied to the product
 of $W_z$.

 Consider to this end the gauge-invariant quantity
 \be
 W_3 (x,y) \equiv tr_Y \prod^3_{i=1} W_{z_i}(x,y)
 \label{7}
 \ee

 We can now write the nonabelian Stokes theorem expressing $A_\mu$
 in $W_{z_i}$ (e.g. using  a general contour gauge) through
 $F_{\mu\nu}(u,x^{(0)})= \phi(x^{(0)}, u) F_{\mu\nu} (u) \phi(u,
 x^{(0)})$ where $\phi(x,y)$ is a parallel transporter, and it is
  convenient to choose $x^{(0)}$ at the common point
 $x=x^{(1)}=x^{(2)} =x^{(3)}$.

 Making final points also coincident, $y=y^{(1)}=y^{(2)}=y^{(3)}$,
 one can use the identity
\be
tr_Y \phi_{a_1b_1} (x,y)  \phi_{a_2b_2} (x,y) \phi_{a_3b_3}
(x,y)=1 \label{8} \ee and rewrite  $W_3$ as (the simplest way to
derive (\ref{9}) and subsequent equations is to choose in the gauge
invariant expressions (\ref{7}), (\ref{9}) the contour  gauge,
where $\phi_{ab} =\delta_{ab}$)
\be
\lan W_3(x,y)\ran = tr_Y \exp \sum^\infty_{n=0} \frac{(ig)^n}{n!}
\int_{\sum S_i} \llan F(1)...F(n)\rran ds(1)... ds(n).
\label{9} \ee ,where in $F(i)$ and the surface elements $ds(i)$
 the Lorentz indices and coordinates
are omitted,and double angular brackets imply cumulants,as defined
in \cite{23}. Note that integration in (\ref{9}) is over all three
lobes $S_i$,  made of contours $C_{z_i} (x,y)$ and the string junction
trajectory $z_Y(s)$, with $z_Y(0)=y, z_Y(1) =x$. The actual form
of $z_Y(s)$ is defined by the minimal action principle and
 not necessarily coincides with
 the trajectory of the center of mass of $3q$ system.
 An important specification is needed at this point.

 For this case of 3-lobe loop as well as for several separate
 loops one can use the following gauge-invariant averaging
 formula, where both field correlators are transported to one
 point $x$ and $a,b,c$ are fundamental color indices
 \be
 \lan F(u,x)_{ab}
 F(v,x)_{cd}\ran = \frac{\lan tr(F(u,x)F(v,x))\ran}{N^2_c-1}
 (\delta_{ad} \delta_{bc} -\frac{1}{N_c} \delta_{ab}
 \delta_{cd}).
 \label{10}
 \ee
 Now whenever $F(u,x), F(v,x)$ are on the same lobe, then indices
 $b$ and $c$ coincide and one obtains
 \be
 tr_Y\lan F(u,x)_{ab} F(v,x)_{bd}\ran =
 \frac{\lan tr~ F(u,x)F(v,x)\ran}{N_c}
 \label{11}
 \ee
 where (\ref{8}) was used. For $u,v$ on different lobes,
 one instead has
\be
tr_Y\lan F(u,x)_{ab} F(v,x)_{cd}\ran = \frac{\lan
tr(F(u,x)F(v,x))\ran}{N_c(N_c-1)}. \label{12} \ee As the last step
in this chapter, one can
 include the quark spin operator
 $\sigma_{\mu\nu} F_{\mu\nu}$
  into the cluster expansion (\ref{9}), with the help of relation
$$
 \lan F_{\mu\nu}(u,x) \exp [ig \int_S F_{\lambda\sigma } (v,x)
 ds_{\lambda\sigma} (v)]\ran=
  $$
  \be
  =\frac{1}{ig} \frac{\delta}{\delta s_{\mu\nu} (v)} \lan
  \exp [ig \int_S F_{\lambda\sigma} (v,x) ds_{\lambda\sigma}
  (v)]\ran.
  \label{13}
  \ee

  Exponentiating the operator  $F_{\mu\nu}$
  one arrives at the shift operator $\exp \frac{1}{i}
  (s_{\mu\nu}\frac{\delta}{\delta s_{\mu\nu}(u)})$
  and finally gets (cf. Eq. (\ref{4})
  $$\lan[ W_3\exp g\sum^3_{i=1}\sigma_{\mu\nu}^{(i)}
  \int^{s_i}_0 F_{\mu\nu} (z^{(i)}
  (\tau^{(i)}))d\tau^{(i)}]\ran\equiv
  \lan W_3\exp (g\sigma F)\ran=
  $$
  \be
  tr_Y\exp [ \sum^\infty_{n=0} \frac{(ig)^n}{n!} \int_{\sum
  S_i}\llan F(1)...F(n)\rran d\rho(1)...d\rho(n)]
  \label{14}
  \ee
  where we have defined $d\rho(n)=\sum^3_{i=1} d\rho^{(i)}(n)$
  \be
  d\rho^{(i)}(n) =ds^{(i)}_{\mu_n\nu_n} (u^{(n)})+\frac{1}{i}
  \sigma^{(i)}_{\mu_n\nu_n} d\tau^{(i)}(n).
  \label{15}\ee
  Here index $i=1,2,3$ refers to three lobes $S_i$ of the total surface
  and it is understood that whenever $F(i)$ under the
  cumulant sign $\llan...\rran$ is multiplied by $d\tau^{(i)}$, it
  is taken at the point $z^{(i)}(\tau_n^{(i)})$,
  lying on the quark trajectory $z^{(i)}(\tau)$ which
   forms the boundary of the lobe $(i)$.

Inserting   (\ref{4}), (\ref{14}) into (\ref{2}) one obtains $$
G_{3q} (x,y)=$$ \be tr_L [\Gamma_{out} \prod^3_{i=1} (m_i-\hat
D^{(i)})_R \int^\infty_0 ds_i(Dz^{(i)})_{xy} e^{-K_i}\lan
W_3\exp(g\sigma F)\ran \Gamma_{in}]. \label{16} \ee

Here $tr_L$ is the trace over Lorentz indices, and $(m_i-\hat
D^{(i)})_R$
 is the value of operator $(m_i-\hat D^{(i)})$
when acting on the path integral, which was found in \cite{43}  to
be
\be
(m_i-\hat D^{(i)})_R=m_i-i\hat p^{(i)}. \label{17} \ee

Here $p^{(i)}$ is the operator of momentum of the quark $i$.
Eqs. (\ref{16}), (\ref{14}) give an exact and most general
expression for the $3q$ Green's function, which is however
intractable if all field correlators are retained there.

To simplify we shall use the observation from lattice calculations
\cite{25,26} that lowest (Gaussian)
 correlator gives the dominant contribution (more than
 95\%) to the static $Q\bar Q$ quark potential. Assuming that
 situation is similar for $3Q$ case and also for light
 baryons, we now keep in (\ref{14}) and (\ref{9}) only
 lowest cumulant $\llan FF\rran$ and express it
 in terms of scalar function $D,D_1$ as in \cite{21}
 $$
 \frac{g^2}{N_c} \lan tr F_{\mu\nu}(u,x) F_{\rho\lambda}(v,x)\ran=
$$\be = (\delta_{\mu\rho}\delta_{\nu\lambda}-\delta_{\mu\lambda}
\delta_{\nu\rho}) D(u-v) +\frac12 [\frac{\partial}{\partial u_\mu}
(u-v)_\rho\delta_{\nu\lambda}+ perm.] D_1(u-v).
 \label{18}
  \ee
  Here
we have replaced parallel transporters $\phi(u,x) \phi(x,v)$ by
the straight-line transporter $\phi(u,v)$, since for the generic
situation with $|u-v|\sim T_g,~~|u-x|\sim|v-x|\sim R,~~ R\gg T_g$,
the former and the latter are equal up to the terms
$O((T_g/R)^2)$. Now in view of (\ref{12}), (\ref{13}) one can
write in Gaussian approximation $$ \lan W_3\exp(g\sigma F)\ran=
\exp [-\frac{g^2}{2N_c}\sum^3_{i=1}\int \lan tr F(u) F(v)\ran
d\rho^{(i)}(u) d\rho^{(i)} (v)- $$
\be
-\frac{g^2}{N_c(N_c-1)}\sum_{i\neq j} \int\lan tr~F(u) F(v)\ran
d\rho^{(i)} (u) d\rho^{(j)} (v)] \label{19} \ee where
$d\rho^{(i)}$ is defined in (\ref{15}). Here $\lan tr FF\ran$ can
be expressed in terms of $D,D_1$ and one has a closed expression
for the term $\lan W_3\exp (g\sigma F)\ran$, which acts as a
dynamical kernel in the path integral (\ref{16}).

Now for large sizes of Wilson loop $W_3$, such that $R^{(i)}\gg
T_g$, one can discard $D_1$ and retain $D$ in (\ref{18}), since
only the latter ensures area law (and moreover, lattice data
\cite{30} show that $D_1\ll D$). Then the diagonal terms in the
sum of the exponent in (\ref{19}) can be written
 as (neglecting spin-dependent part for the moment)
 \be
 \lan W_3\ran_{diag} =\exp (-\sigma(S_1+S_2+S_3))\label{20}
 \ee
 where $S_i$ is the area of the minimal surface between trajectory
 of quark $(i)$ and trajectory of string junction (Y-trajectory),
 and we have used the relation \cite{21,22}
 \be
 \sigma=\frac12\int d^2xD(x).
 \label{21}
 \ee

 Let us turn now to nondiagonal terms in (\ref{19}).
 Since $D(x), D_1(x)$ are dying exponentially fast for
 $x>T_g$\cite{30}-\cite{32},
 only  region of width $T_g$ around $Y$ trajectory
 contributes to these terms, which one can write as
 $$
 V_{nondiag.}T=\sum_{i\neq j} V^{(ij}_{nondiag.}=$$\be=
 \frac{g^2}{N_c(N_c-1)} \sum_{i\neq j}
 \int \lan tr F_{\mu\nu} (u) F_{\rho\lambda}(v)\ran
 ds^{(i)}_{\mu\nu}(u) ds_{\rho\lambda}^{(j)}(v).
 \label{22}
 \ee
 Separating out time components, $u=(u_4,\veu),~~v=(v_4,\vev),$
 one can write for the $D$ contribution
 $$
 V_{nondiag.}^{(D)} =$$\be=\frac{1}{N_c-1}\sum_{i\neq j}\int
 D(u^{(i)}-v^{(j)}) d(u^{(i)}_{\parallel}-v^{(j)}_{\parallel})
 d\left (\frac{u^{(i)}_{\perp}+v^{(j)}_{\perp}}{2}\right)
 d(u^{(i)}_4-v^{(j)}_4)
 \label{23}
 \ee
 where we have introduced for $\veu, \vev$  parallel and
 transverse components on the lobe $S_i$ with respect to the lobe $S_j$.

 Since $|u^{(i)}-v^{(j)}|=\sqrt{(u^{(i)}_4-v^{(j)}_4)^2+
 (u^{(i)}_{\parallel}-u^{(j)}_{\parallel})^2+(u_{\perp}^{(i)}-u_{\perp}^{(j)})^2}$
 is growing fast with $|u_{\perp}-v_{\perp}|$ one can estimate
 (\ref{20}), (\ref{23}) as
 \be
 V^{(D)}_{nondiag.}\sim \sigma T_g,~~
 V^{(D)}_{diag.}=\sigma(r^{(1)}+r^{(2)}+r^{(3)}).
 \label{24}
 \ee

 where $\ver^{(i)}=\vez^{(i)}-\vez^{(Y)}$ , i.e. the difference of quark and
string junction coordinates at a given moment of time.
 Being always smaller than $V_{diag.}^{(D)}$ for large $r^{(i)}$,
 nevertheless $V^{(D)}_{nondiag.}   $ brings about an interesting
 cancellation for small $r^{(i)}$. Estimating integrals in
 (\ref{23}) for small $r^{(i)}$, $r^{(i)} \la T_g$,
 and using for $D(x)$ the Gaussian form,
 $D(x)=D(0) \exp (-\frac{x^2}{4T_g^2})$, one has
 \be
 V_{conf}=V_{diag.} +V_{nondiag.} = \frac{\sigma}{2\sqrt{\pi} T_g}
 (\sum_i \veR^{(i)})^2.
 \label{25}
 \ee

 It is clear that for a symmetric configuration
 $r^{(1)}=r^{(2)}=r^{(3)}$ one has $V_{conf}=0.$
 To study further this cancellation let us take into
 account that if the triangle made of quarks has all
 angles less than 120$^o$ (since string junction is at Torricelli
 point)then the string junction is inside this triangle. In this
 case one can write
 $$
 \sum_{i<j}(\ver^{(i)}-\ver^{(j)})^2=$$\be =
 2((r^{(1)})^2+(r^{(2)})^2+(r^{(3)})^2)+r^{(1)}r^{(2)}
 +r^{(2)}r^{(3)}+r^{(1)}r^{(3)}
 \label{26}
 \ee
 i.e., $V_{conf}$ vanishes quadratically in differences of
 quark distances from the string junction.
Practically this brings about a strong effective cancellation in
$V_{conf}$
 for $3q$ system with equal masses
at approximately equal distances. Numerically and analytically
 this fact was discovered first in \cite{44}
for static $3Q$ potential. It was argued there that $V_{nondiag}$
brings about smaller slope of $V(3Q)$
 at small to intermediate distances, as was indeed found on the lattice
 \cite{45}.
An assumption that a triangular $3q$ string configuration is
responsible for the smaller slope, however, cannot explain it,
since that configuration  is impossible to construct in a
gauge-invariant way \cite{46}. Explicit expressions for $V^{(3)}$
in general case are given in  \cite{46}. Here and in \cite{46}
a missing in \cite{44} factor of $-1/2$ in front of $V_{nondiag}$
is restored.

\section{Gaussian representation for the effective action
of quarks and string}

Consider now the exponent of the FFS representation for the $3q$ Green's function
(\ref{16}), (\ref{19}) in the simplified case when  i) spin
interaction is neglected and ii) large distances $|\veR_i|\gg T_g$
are taken into account.

In this situation one can use the form (\ref{20}) instead of
(\ref{19}), and writing the exponential term in (\ref{16}) as
\be
G_{3q} (x,y) =tr_L[\Gamma_{out}\prod(m_i-i\hat{p}_i) \int^\infty_0
ds_i(Dz^{(i)}_4)_{xy} \Gamma_{in}]e^{-A} \label{27} \ee where $A$
plays the role of effective action,
\be
A=\sum^3_{i=1} (K_i+\sigma S_i). \label{28} \ee Our purpose is
finally to construct the effective Hamiltonian,
 considering $A$ as an effective
action for 3 quarks and the composite string with the string
junction. To achieve this goal one must i) go over from the proper
time $s_i$ to the real time integration in the $4d$
Euclidean space-time (later on to be transformed into Minkowskian time
), ii) to transform the Nambu-Goto form of the lobe area $S_i$
(see below in (\ref{35})) into a quadratic
form, as it is necessarily done in string theory (since otherwise
the path integral (\ref{27}) is not properly defined). Both
operations are the same as in the $q\bar q$ case, considered in
\cite{34} and we shall follow closely that procedure.

The resulting Hamiltonian depends on the choice of the
hypersurface, and for the $q\bar q$ system both the c.m. \cite{34}
and light-cone \cite{47,48} cases were considered.

Below  in the next chapter the c.m. Hamiltonian will be derived,
and to this end we choose
 the hyperplane intersecting all
 3 quark trajectories and $Y$ trajectory at one common time $t$,
 to be considered in the interval $0\leq t\leq T$, so that
 quark coordinates are $z^{(i)}= (t,\vez^{(i)})$, and string
 junction coordinate is $z^{(Y)}=(t,\vez^{(Y)})$.

 Now one can make a change of variables, introducing the einbein
 variable \cite{34,49} $\mu(t)$ for a given trajectory
 $z^{(i)}(\tau^{(i)})$, $0\leq\tau \leq s$, one defines
 \be
 d\tau^{(i)} \equiv \frac{dz^{(i)}_4(\tau^{(i)})}{\frac{dz^{(i)}_4(\tau^{(i)})}
 {d\tau^{(i)}}}=\frac{dz_4^{(i)}(\tau^{(i)})}{2\mu_i(z_4^{(i)})}
 \label{28a}
 \ee
so that kinetic term $K_i$ becomes
\be
K_i= \int^T_0 dt [\frac{m^2_i}{2\mu_i(t)} + \frac{\mu_i(t)}{2}
(\dot{\vez}^2(t)+1)]. \label{29} \ee The transition from the
integral over $ds_i dz^{(i)}_4$ to the integral over
$D\mu^{(i)}(t)$ is known to have a nonsingular Jacobian (see
Appendix A of the second paper in \cite{34}
   for more details and explanations)
   \be
   D\mu^2(t)\sim\exp [-i\frac{const}{\varepsilon}
   \int^T_0\sqrt{\mu^2(t)} dt] ds Dz_4 (t)\label{30}
   \ee
   where $\varepsilon \sim 1/\Lambda$, and $\Lambda$ is an
   ultraviolet cut-off parameter.

    Hence the integrals in (\ref{27}) can be rewritten as
    \be
    \prod_i ds_i D^4z^{(i)}\to \prod D\mu_i D^3 z^{(i)}\label{31}
    \ee
    where the integration measure for $D\mu_i$ can be specified
    further to be \cite{34}: $D\mu (t) \sim
    \prod^N_{n=1}\frac{d\mu(t_n)}{\mu^{3/2}(t_n)}$.
    As a next step one introduces the c.m. and relative
    coordinates
$$
   \dot{R}=\frac{1}{\mu_+(t)} \sum^3_{i=1}\mu_i(t) \dot{\vez}^{(i)} (t)
  $$
\be
    \dot{\vexi} =\sqrt{\frac32}(\frac{\mu_1\dot{\vez}^{(1)}+\mu_2\dot
    {\vez}^{(2)}}{2} -\mu_3\dot{\vez}^{(3)})\frac{1}{\mu_+},~~
    \dot{\veta}=\frac{\mu_1\dot{\vez}^{(1)}-\mu_2\dot
{\vez}^{(2}}{\mu_{+}\sqrt{2}}.
    \label{32}
    \ee
 Here $\mu_+=\sum^3_{i=1} \mu_i$.

    From our discussion above it is clear that the time $t$
    coincides with the fourth component of the c.m. coordinate,
    $t=R_4$, and the whole quark kinetic term in (\ref{28}) is:

    \be
    \sum^3_{i=1} K_i = \int^T_0 dt \left[\sum\left(
    \frac{m^2_i}{2\mu_i}+\frac{\mu_i}{2}\right)+\frac12 \mu_+ (t)
    \dot{ \veR^2}+\frac12\mu_\eta
    \dot{\veta^2}+\frac12\mu_\xi\dot{
    \vexi^2}\right].
    \label{33}
    \ee

    Here $\mu_\eta$ and $\mu_\xi$ will be found below, (\ref{42}).
    The area-law term in (\ref{28}) can be written as follows
    \be
    \sum^3_{i=1}\sigma S_i=\sigma \sum^3_{i=1} \int^T_0 dt
    \int^1_0 d\beta_i \sqrt{(\dot
    w^{(i)}_\mu)^2(w^{'(i)})^2-(\dot w_\mu^{(i)}
    w^{'(i)}_\mu)^2}\label{34}
    \ee
    where $w_\mu^{(i)}(t,\beta)$ is the $i$-th string position at
    the time $t$ and coordinate $\beta$ along the string and the dot
     and prime signs have the meaning of the time and $\beta$ derivatives
      respectively. In the
    spirit of our approach one should take the world sheets of the
    strings corresponding to the minimal area of the sum of
    surfaces between quark trajectories an $Y$- trajectory of the
    string junction. At this point we make a simplifying
    approximation \cite{33,34} that strings at any moment $t$ can be
    represented by pieces of straight lines. In this way one
    disregards string excitations (hybrids) and mixing between
    these excitations and ground state baryons. This can be done
    for ground states since the mass gap for string excitations is
    around 1 GeV \cite{24}.

    For higher excited states the mixing should be taken into
    account analogously to what was done in meson sector \cite{50}.

    Thus one writes
    \be
    w^{(i)}_\mu (t,\beta) =z^{(i)}_\mu(t) \beta+ z^{(Y)}_\mu(t)
    (1-\beta)\label{35}
    \ee
    and time derivatives of $w_\mu^{(i)}$ in (\ref{34}) can be
    replaced using (\ref{35}) by time derivatives of $z^{(i)}_\mu,
    z_\mu^{(Y)}$. Since also the string junction position is expressed
    through the quark coordinates, the string does not possess the
    dynamical degrees of freedom of its own (in this straight-line
    approximation). To recover the latter one can use background
    perturbation theory and consider the states with $3q$ and
    additional valence gluon(s). The latter describes gluonic
    excitation of baryon and has its own dynamical degree of
    freedom. Note that this way of systematic description of
    string excitation  is different from the $ad~  hoc$ assumption
    that string is described by Nambu-Goto action with all
    dynamical string degrees of freedom included,which does not follow
    from the QCD Lagrangian.

    Consider now string-junction trajectory. In line with the
    whole approach one requires that at any given moment
    $z^{(Y)}_\mu(t)$ occupies the position which gives the minimal
    string energy, i.e. $z^{(Y)}_\mu(t)$ should coincide with the
    Torricelli point, giving the minimum of the sum of lengths of
    3 strings:
    \be
    L=\sum^3_{i=1} |\vez^{(i)} (t) -\vez^{(Y)} (t)|, ~~
    \frac{\partial L}{\partial z^{(Y)}_k(t)} =0.
    \label{36}
    \ee

    Therefore $\vez^{(Y)}(t)$ is not an independent  dynamical
    degree of freedom and $\dot{ \vez}^{(Y)}$ is expressed in terms of
    $\dot{    \vez}^{(i)},~i=1,2,3.$

    Now one can introduce (as is usual in string theory \cite{51})
    the auxiliary fields (einbein fields \cite{49}) to replace the
    untractable square-root terms in (\ref{34}) by quadratic
    expressions. In this way one writes
    \be
    S_i=\frac{1}{2\tilde \nu_i}[(\dot w^{(i)})^2+ (\sigma\tilde
    \nu_i)^2(r^{(i)})^2-2\eta_i (\dot w_k^{(i)}r_k^{(i)})+
    (\eta_i)^2(r^{(i)})^2].\label{37}
    \ee

    Here $\tilde\nu_i (t,\beta)\geq 0$ and $\eta_i(t,\beta)$ are two
    einbein fields( which are integrated out to yield back the
    form (\ref{34})), and $\ver^{(i)} =\vez^{(i)}-\vez^{(Y)}$. As a
    result one has for the $3q$ Green's function
    \be
    G_{3q} (x,y) = \int DRD\xi D\eta \prod^3_{i=1} D\mu_i D\tilde
    \nu_i D\eta_i tr(\Gamma_{out}
    (m_i-i\hat{p_i})\Gamma_{in})e^{-A}.\label{38}
    \ee

    \section{Quantization of the strings and derivation of the
    $3q$-string Hamiltonian}

\newcommand{\vel}{\mbox{\boldmath${\rm l}$}}

    The action (\ref{28}) using (\ref{33}), (\ref{35}), (\ref{37})
     and can be written
    as
    $$
    A=\int^t_0 dt \sum^3_{i=1} \left [
    \frac{m_i^2}{2\mu_i}+\frac{\mu_i\dot{ \vez}^2_i}{2}
    +\frac{\mu_i}{2} +\int^1_0 d\beta_i\frac{\sigma^2
    r^2_i}{2\nu_i}+\right.
    $$
    $$
    +\frac12 \int^1_0 d\beta_i\nu_i
    (\dot{\ver}_i\beta_i+\dot{\vez}_Y)^2 + \frac12 \int^1_0
    d\beta_i\nu_i \eta_i^2\ver^2_i-
    $$
    \be
\left.    -\int^1_0 d\beta_i\nu_i\eta_i\ver_i (
    \dot{\ver}_i\beta_i+\dot{\vez}^{(Y)})\right]
    \label{40}     \ee
 where we have defined $\nu_i=1/\tilde \nu_i$, and
$\ver_i = \vez^{(i)}- \vez^{(Y)},~~ z^{(Y)}_\mu(t) =(t, \vez^{(Y)}).$

As the next step we introduce the c.m. coordinate $\veR$ and
Jacobi coordinates $\vexi, \veta$ as follows \cite{36} $$ \dot
{z}_k^{(1)} = \dot {R}_k+ \left (
\frac{\mu\mu_3}{\mu_+(\mu_1+\mu_2)}\right)^{1/2} \dot \xi_k
-\left(
\frac{\mu\mu_2}{\mu_1(\mu_1+\mu_2)}\right)^{1/2}\dot{\eta}_k $$ $$
\dot {z}_k^{(2)} = \dot {R}_k+ \left (
\frac{\mu\mu_3}{\mu_+(\mu_1+\mu_2)}\right)^{1/2} \dot \xi_k+\left(
\frac{\mu\mu_1}{\mu_2(\mu_1+\mu_2)}\right)^{1/2}\dot{\eta}_k $$
\be
\dot {z}_k^{(3)} = \dot {R}_k- \left (
\frac{\mu(\mu_1+\mu_2)}{\mu_+\mu_3}\right)^{1/2} \dot \xi_k,
\label{41} \ee with the inverse  expressions $$ \dot
R_k=\frac{1}{\mu_+}\sum^3_{i=1} \mu_i \dot z^{(i)}_k,~~ \dot{
\eta}_k= (\dot z^{(2)}_k-\dot z^{(1)}_k) \left (
\frac{\mu_1\mu_2}{\mu(\mu_1+\mu_2)}\right)^{1/2} $$
\be
\dot{\xi}_k=\left (
\frac{\mu_3}{\mu_+(\mu_1+\mu_2)\mu}\right)^{1/2} (\mu_1\dot
z^{(1)}_k+\mu_2\dot z_k^{(2)}- (\mu_1+\mu_2)\dot z^{(3)}_k)
\label{42}\ee

 In (\ref{41}), (\ref{42}) the mass $\mu$ is arbitrary and drops
 out in final expressions.

 Using (\ref{41}) one can rewrite the kinetic part of the action
 as follows
 \be
 \sum^3_{i=1} K_i=\int^T_0 dt [\sum^3_{i=1} \left(
 \frac{m^2_i}{2\mu_i}+\frac{\mu_i}{2}\right) +
 \frac12\mu_+\dot{ \veR}^2+\frac12\mu
 (\dot{\veta}^2+\dot{\vexi}^2)].
 \label{43}
 \ee

The string part of the action can be transformed using (\ref{37})
and integrating over $\eta_i$ to the form $$ \exp
(-\sum^3_{i=1}\sigma S_i) = \exp \{-\int^T_0 dt \frac12
\sum^3_{i=1} \int d\beta_i [ \nu_i ( (\dot {\ver}^{(i)}\beta_i+
\dot{ \vez}^{(Y)})^2-
  $$
  \be
  -[(\dot{\ver}^{(i)} \beta +\dot{\vez}^{(Y)})
  \ver^{(i)}]^2\frac{1}{(\ver^{(i)})^2})+\frac{\sigma^2\ver^2_i}{\nu_i}
  ]\}.
  \label{44}
  \ee
  At this point it is important to note that $\vez^{(Y)}$ is not a
  dynamical  variable, since it is defined by the minimum of the
  action.

  Taking this minimum at a given moment, one arrives at the
  definition of $\vez^{(Y)}(t)$ as a Torricelli point, which is to
  be expressed through the  positions $\vez^{(i)}(t)$;
  \be
  \vez^{(Y)} (t) = f(\vez^{1)}(t), \vez^{(2)} (t),
  \vez^{(3)}(t))\label{45}
  \ee
  where the function $f$ is defined explicitly in \cite{36}.
   Therefore $\dot {\vez}^{(Y)}(t)$ is also expressed  through
  $\dot {\vez}^{(i)}(t)$, or through $\dot {\veR}(t)$ and
  $\dot{\ver}^{(i)}(t)$.

Below the simplified procedure will be used, where one identifies
$\vez^{(Y)}$ with the c.m. coordinate $\veR$, which is true on
average for equal mass quarks. Explicit formulas for the general
case $\vez^{(Y)}\neq \veR$
 are given in Appendix. We are now in position to
get the final coordinates $\veta, \vexi$ or their linear
combinations $\ver^{(i)}\equiv \vez^{(i)}-\veR$ read from
(\ref{41}).
 To this end we replace in (\ref{44}) $\dot {\vez}^{(Y)}$
by $\dot{\veR}$ and integrate over
 $D\dot{\veR}$ in both expressions (\ref{43}), (\ref{44}) in
the same way, as was done in the second paper of( \cite{34}, Eqs.
(\ref{36})-(\ref{49})), with the result $$ \tilde A =\int^T_0 dt
\frac12 \left\{\sum^3_{i=1} \left(\frac{m^2_i}{\mu_i}+ \mu_i
\right)+\mu(\dot{\veta}^2+\dot{\vexi}^2)+\sum^3_{i=1} \int^1_0
d\beta_i[\nu_i(\beta_i) +\right. $$
\be
\left.+\frac{\sigma^2(\ver^{(i)})^2}{\nu_i}
 +\nu_i\beta^2_i ((\dot {\ver}^{(i)})^2
 -\frac{(\dot{\ver}^{(i)}\ver^{(i)})^2}{(\ver^{(i)})^2})]\right\}
.
\label{46} \ee
 The last term on the r.h.s. of (\ref{46}) can be rewritten as
 $\nu_i\beta_i^2\frac{(\dot{\ver}^{(i)}\times\ver^{(i)})^2}{\ver^2}$
 and disappears when the partial angular momentum $l_i$ vanishes.

 In this case (\ref{46}) simplifies and using in (\ref{46})
 $\dot{\ver}^{(i)}$ instead of $\dot{\veta},\dot{\vexi}$, one gets
 in a standard way the Hamiltonian
 \be
 H_0=\frac12\sum^3_{i=1}
 [\frac{m^2_i}{\mu_i}+\mu_i+\frac{\vep^2_i}{\mu_i} +\int^1_0
 d\beta_i(\nu_i(\beta_i)+\frac{\sigma^2(\ver^{(i)})^2}{\nu_i})].
 \label{47}\ee
 One can now apply to (\ref{47}) the minimization procedure to
 define $\mu_i,\nu_i$ from the conditions
 \be
 0=\frac{\partial H_0}{\partial\mu_i} =
  \frac{\partial H_0}{\partial\nu_i}
  \label{48}
  \ee
  which yields
  \be
  \mu_i=\sqrt{\vep^2_i+m^2_i}, ~~\nu_i=\sigma|\ver_i|.
  \label{49}\ee
  Note that in this case $(l_i=0,~~i=1,2,3) \nu_i$  do not
  depend on $\beta_i$ and play the role of  potential. Inserting
  (\ref{49}) in (\ref{47}) one obtains the form well known from
  the standard relativistic quark model (RQM)\cite{1}-\cite{9}
\be
H_{RQM} = \sum^3_{i=1} (\sqrt{\vep^2_i+m^2_i}+\sigma|\ver^{(i)}|).
\label{50} \ee

Note that (\ref{50}) is valid under assumptions that (1)
 string-junction $\vez^{(Y)}$ coincides
with the c.m.; (2) $\sum^3_{i=1}\vep_i=0$; (3) all angular momenta
of quarks $l_i,i=1,2,3$ are zero, so that only radial part of
momentum $\vep_i$ enters in (\ref{50}). However, in RQM the form
(\ref{50}) is used without the condition (3). As one will see in
what follows, at nonzero $l_i$ the Hamiltonian $H_0$ will be
modified, and for not large $l_i, l_i \leq 4$ this  modification
 can be taken into account as a string correction $\Delta H_{string}$
similarly to the meson case in \cite{34,27}.

We consider now the general case of $l_i>0$. To this end we
separate for each $\ver^{(i)}$ transverse and
 longitudinal components as follows
(omitting index $i$ for a moment)
\be
\dot{\ver}^2=\frac{1}{\ver^2}
\{(\ver\dot{\ver})^2+(\dot{\ver}\times \ver)^2\}\label{51} \ee and
correspondingly define  transverse and longitudinal momenta
\be
p^2_r=\frac{(\vep\ver)^2}{r^2}=\frac{(\mu\ver\dot{\ver})^2}{r^2},
\label{52} \ee
\be
p^2_T=\frac{(\vep\times\ver)^2}{r^2}= (\mu+\int_0
d\beta\beta^2\nu(\beta))^2\frac{(\dot{\ver} \times \ver)^2}{r^2}.
\label{53} \ee One can now derive the Hamiltonian from (\ref{46})
in  the usual way $$ H=\sum^3_{i=1}
[\frac{m^2_i+p_{ri}^2}{2\mu_i}+\frac{\mu_i}{2}+ \frac{\hat
l^2_i/r^2_i}{2(\mu_i+ \int^1_0 d\beta_i\beta_i^2\nu_i(\beta))}+ $$
\be
+\frac{\sigma^2}{2}\int^1_0 \frac{d\beta_i}{\nu_i(\beta_i)}
\ver^2_i+\frac12 \int^1_0\nu_i(\beta_i)d\beta_i] \label{54} \ee

This is a general form for any values of $l_i$, the limit of
$l_i\to 0$
 is obtained in (\ref{50}).
Now we shall derive the opposite  limit $l_i\to\infty$. As in the
meson case one can argue that in this case $\mu_i\ll \nu_i$ and
one can use
 the quasiclassical method and retain
in (\ref{54}) only the last three terms, expanding them around the
stationary point at $r^{(i)}=r^{(i)}_0$, where
\be
(r^{(i)}_0)^2=\left[ \frac{\hat l^2_i}{2(\mu_i+ \int^1_0
d\beta_i\beta_i^2\nu_i(\beta))
\sigma^2\int^1_0\frac{d\beta_i}{\nu_i(\beta_i)}}\right]^{1/2}.
\label{55} \ee

Inserting (\ref{55}) back into (\ref{54}) one obtains the
following quasiclassical energy of the $i-$th string,
$E=\sum^3_{i=1}  E_i$
\be
E_i=\sigma\sqrt{\hat l^2_i}\left(\frac{\int^1_0
d\beta_i\nu_i^{-1}}{ \int^1_0 d\beta_i
 \beta^2_i \nu_i}\right)^{1/2} +\frac12
 \int^1_0 \nu_i d\beta_i
 \label{56}
 \ee
  and from the stationary point of $E_i,~~
\frac{\delta E_i}{\delta \nu_i(\beta_i)}=0,$
 one has
 \be
 \nu_i(\beta_i) =\sqrt{\frac{2}{\pi}}(1-\beta_i^2)^{-1/2}.
 \label{57}
 \ee
 Inserting (\ref{57}) into (\ref{56}) one obtains finally the
 energy of rotating string
 \be
 E^2_i= 2 \pi \sigma \sqrt{\hat l^2_i},~~ \hat l^2_i \equiv
 l_i(l_i+1).
 \label{58}
 \ee
 This result shows that our general baryonic Hamiltonian indeed
 admits simple rotating string limit at large $l_i$, as it was in
 the case of mesonic Hamiltonian.

 The difference from the mesonic case is however, that for the
 $3q$ system one should be careful in proper exclusion of the c.m.
 motion and in quantizing angular momenta $l_i$, which should add
 up to a total angular momentum $\veL=\sum^3_{i=1} \vel_i.$ To do
 this one should go over from $\ver^{(i)}, l_i$ to  the
 independent Jacobi coordinates and momenta $\vexi, \veta,
 \vel_\xi, \vel_\eta$ which add up to $\veL$ as
 $\veL=\vel_\xi+\vel_\eta$.

 To accomplish this task, one should express $r_k^{(i)}=
 z^{(i)}_k-z_k^{(Y)}$ using (\ref{41})  and Appendix through $\vexi, \veta$ and
$ \vel_i$ through $\vel_\xi, \vel_\eta$, and insert it into
(\ref{54}) which
 makes the whole expression rather complicated and not very
 tractable.
 Instead we adopt here another strategy and consider the
 contribution of string rotation as a correction, similarly to the
 case of mesons  \cite{34}, where this approach has proved to be
 successful up to $L=4$ \cite{27}.  Therefore we shall represent the
 Hamiltonian (\ref{54}) as a sum of unperturbed term $H_0$ plus a
 string correction $\Delta H_{string}$, which should work with
 accuracy better than 5\% up to $l\approx 3\div 4$,
 \be
 H=H_0+\Delta H_{string}
 \label{59}\ee
 where we have defined
 \be
 H_0=\sum^3_{i=1}
 \left(\frac{m^2_i}{2\mu_i}+\frac{\mu_i}{2}\right)
 +\frac{\vep^2_\xi+\vep^2_\eta}{2\mu} +V_{conf.}(\ver_1,\ver_2,
 \ver_3)
 \label{60}
 \ee
 and
 \be
 V_{conf.}= \sigma\sum^3_{i=1} |\vez^{(i)}- \vez^{(Y)}| =\sigma
 \sum^3_{i=1}r_i
 , \label{61}\ee

 The string correction $\Delta H_{string}$ is computed
  to be
    \be
  \Delta H_{string}=-\sum^3_{i=1} \frac{\hat l^2_i \sigma <
  r^{-1}_i>}{2\lan \sigma r_i\ran (\mu_i+\frac13<\sigma r_i>)}.
  \label{62}
  \ee

  The total Hamiltonian for the bound 3q system in the c.m.
  coordinates, taking into account only valence quarks, can now be
  written as follows.
  \be
  H_{tot} =H_0+\Delta H_{string} + \Delta H_{coul.} +\Delta
  H_{self.}+\Delta H_{spin}
  \label{63}
  \ee
  where $H_0$ is given in (\ref{60}), $\Delta H_{string}$ in
  (\ref{62}), $\Delta H_{spin}$ is given in
  \cite{40}, $\Delta H_{coul.}$ is easily computed allowing for
  perturbative gluon exchanges in $W_3$, resulting in a standard
  expression
  \be
  \Delta H_{coul.} =- \frac{2\alpha_s}{3} \sum_{i<j}
  \frac{1}{|\vez^{(i)}-\vez^{(j)}|}.\label{64}
  \ee
  As to $\Delta H_{self.} $, it was found in \cite{41}  to originate
  from the $\lan \sigma F\sigma F\ran$ correlator referring to the
  same quark line. It has the form (\cite{41}, Eq.(\ref{35}))
  \be
  \Delta H_{self} = -\frac{2\sigma}{
  \pi} \sum^3_{i=1}
  \frac{\eta_i}{\mu_i}.
  \ee
where $\eta_i=1$ for light quarks.

\section{The light-cone quantization of the $3q$ system.
Derivation of the light-cone Hamiltonian}

The general expression of the 3q Green's function allows to
calculate Hamiltonian, corresponding
 to any prescribed hypersurface, with the evolution parameter $T$
 orthogonal to it, according to the equation (in Euclidean
 space-time)
     \be
     \frac{\partial G}{\partial T} =-HG.
     \label{5.1}
     \ee

In the previous section the hypersurface was chosen to be
$z_4^{(i)} = const.$, and  the corresponding c.m. Hamiltonian  was
written in  (\ref{63}). The  obtained  Hamiltonian is a
3q equivalent of the $q\bar q $ c.m. Hamiltonian  found earlier in
\cite{34,47,48}.

The light-cone version of the $q\bar q$ Hamiltonian was derived in
\cite{47} and solved numerically
 in \cite{48}.

 In this section we shall follow the same technic as in \cite{47}
  to obtain the 3q Hamiltonian on the light cone. To this end one
  should choose the hypersurface to be the plane with fixed values
  of $z^{(i)}_+$, where we use the following convention
  \be
  ab=a_\mu b_\mu = a_i b_i - a_0 b_0 =a_\bot b_\bot+ a_+b_- + a_-
  b_+
  \label{5.2}
  \ee
  and $a_{\pm} = \frac{a_3\pm a_0}{\sqrt{2}}$.

  The same decomposition of  quark coordinate $z_\mu^{(i)}$ will
  be used as in (\ref{41}), but for the light cone (l.c.)
  coordinates (\ref{5.2}). Again for simplicity we shall identify
  $R_\mu$ and $z_\mu^{(Y)}$, so that
  \be
  r^{(i)}_\mu=z^{(i)}_\mu- R_\mu= z_\mu^{(i)} -
  z_\mu^{(Y)}\label{5.3}
  \ee

  Some kinematical properties of l.c. coordinate to be used below
  are
 $$
 w^{(i)}_\mu
 (z_+,\beta_i)=z^{(i)}_\mu\beta_i+z^{(Y)}_\mu(1-\beta_i)=
 $$
 \be
 =r^{(i)}_\mu\beta_i+z_\mu^{(Y)}\cong r^{(i)}_\mu\beta_i + R_\mu,
 \label{5.4}
 \ee
 $$
 r^2_\mu = r^2_\bot,~~ r_+\equiv 0;~~ \frac{\partial
 w^{(i)}_\mu}{\partial \beta_i} = r_\mu^{(i)};
 $$
 $$\dot w_\mu^2=\dot w^2_\bot + 2\dot w_-; ~~ \dot w_\bot = \dot
 r_\bot \beta+\dot R.
 $$

Having this in mind one can directly obtain the l.c. action from
(\ref{40}) (cf \cite{47} for the equivalent derivation of $q\bar
q$ l.c. action) $$ A_{lc}= \int^T_0 dz_+ \sum^3_{i=1} \left \{
\frac{m^2_i}{2\mu_i} +\frac{\mu_i}{2} ((\dot R_\bot +\dot
r_\bot^{(i)})^2+\right. $$ $$ +2(\dot R_- +\dot
r_-(i)))+\frac{1}{2} \int^1_0 d\beta_i \left
[\frac{\sigma^2(r_\bot^{(i)})^2}{\nu_i}+\right. $$ $$ +\nu_i ((
\dot r_\bot^{(i)} \beta_i+\dot R_\bot)^2+ 2 (\dot r^{(i)}_-
\beta_i+\dot R_-)) + \nu_i\eta_i^2r^{2(i)}_\bot- $$ \be
\left.\left. -2\nu_i\eta_i((\dot r_\bot^{(i)}\beta_i+ \dot
R_\bot)r_\bot^{(i)} + r_-^{(i)})\right]\right\}, \label{5.5} \ee
where $\dot r_\mu^{(i)}, (\mu=\perp,-)$ can be expressed in terms
of $\dot \xi, \dot \eta$ using parametrization
\be
\dot R_\mu=\sum^3_{i=1} x_i\dot z_\mu^{(i)}, ~~\sum^3_{i=1} x_i=1, ~~
x_i\geq 0 \label{5.6} \ee

$$ \dot z_\mu^{(1)}-\dot R_\mu=\left(
\frac{x_3}{x_1+x_2}\right)^{1/2}\dot \xi_\mu - \left(
\frac{x_2}{x_1(x_1+x_2)}\right)^{1/2}\dot \eta_\mu $$
\be
\dot z_\mu^{(2)}-\dot R_\mu=\left(
\frac{x_3}{x_1+x_2}\right)^{1/2}\dot \xi_\mu + \left(
\frac{x_1}{x_2(x_1+x_2)}\right)^{1/2}\dot \eta_\mu \label{5.7} \ee
$$ \dot z^{(3)}_\mu-\dot R_\mu =- \left(
\frac{x_1+x_2}{x_3}\right)^{1/2} \dot \xi_\mu. $$

One can rewrite (\ref{5.5}) in the same form as in \cite{47} $$
A_{lc} =\frac12 \int^T_0 dz_+\{ a_1\dot R^2_\bot + 2a_1\dot R_- +
2a_{2\bot} \dot R_\bot + 2a_{2-}- $$ $$ -2c_{1\bot} \dot R_\bot
+\sum^3_{i=1} [- 2c_{2i} \dot r^{(i)}_\bot r^{(i)}_\bot + a_{3i}
(\dot r_\bot^{(i)} )^2- $$
\be
-2c_{1i} r^{(i)}_- + a_{4i}
(r_\bot^{(i)})^2+\frac{m^2_i}{\mu_i}]\} \label{5.8} \ee where we
have defined $$ a_1=\sum^3_{i=1} (\mu_i+\int^1_0\nu_i(\beta)
d\beta)=\sum^3_{i=1} a_{1i} $$ $$ a_{2\bot}=\sum^3_{i=1}
(\mu_i+\int^1_0\nu_i(\beta)\beta  d\beta)\dot r_\bot^{(i)} $$ $$
a_{2-}=\sum^3_{i=1} (\mu_i+\int^1_0\nu_i(\beta)\beta d\beta)\dot
r_-^{(i)} =\sum^3_{i=1} a_{2i} \dot r_-^{(i)}$$
\be
c_{1\bot} = \sum^3_{i=1} \int^1_0 d\beta \nu_i (\beta) \eta_i
r_\bot^{(i)}, ~~ c_{1i} =\int^1_0 \nu_i(\beta) \eta_i d\beta
\label{5.9} \ee $$ c_{2i} =\int^1_0 d\beta \nu_i(\beta) \beta
\eta_i,~~ a_{3i}=\mu_i+\int^1_0\nu_i(\beta) \beta^2 d\beta; $$
\be
a_{4i} =\int^1_0 d\beta (\nu_i \eta_i^2+ \frac{\sigma^2}{\nu_i}).
\label{5.9a} \ee We now require as in \cite{47} that transverse
velocity should be diagonalized, i.e. the mixed term $a_{2\bot}$
to vanish. This gives conditions on coefficients $x_i$, when
$a_{2\bot} $ is expressed in terms of two independent velocities:$
\dot \xi_\bot$ and $\dot \eta_\bot$.

This immediately yields expressions for $x_i$:
\be
x_i=\frac{\mu_i+\int^1_0\nu_i(\beta)
 \beta d\beta}{\sum^3_{i=1} (\mu_i+ \int^1_0 \nu_i(\beta)
\beta d\beta)}. \label{5.10} \ee

  Now one can integrate over $\prod^3_{i=1} D\eta_i$ in the same
  way, as it was done for the $q\bar q$ system in \cite{47} with the
  result
  $$
  A'_{lc} =\frac12 \int^T_0 dz_+\{ a_1 (\dot R^2_\bot +2R_-)
  +\sum^3_{i=1} [ \int^1_0 d\beta \frac{\sigma^2 (r^{(i)}_\bot)^2}{\nu_i}
+
  \frac{m^2_i}{\mu_i} + a_{3i} (\dot r_\bot^{(i)})^2-
  $$
  \be
  -\frac{(r^{(i)}_-+r^{(i)}_\bot\dot R_\bot+
  \lan \beta\ran_i \dot r_\bot^{(i)}
  r_\bot^{(i)})^2\int\nu_i d\beta}{(r^{(i)}_\bot)^2}-
  \frac{(\dot r_\bot^{(i)}r_\bot^{(i)})^2}{(r_\bot^{(i)})^2}
  \int^1_0 \nu_i (\beta) (\beta-\lan \beta\ran_i)^2]\}.
  \label{5.11}
  \ee
  Here we have defined
  \be
  \lan \beta\ran_i = \int^1_0\nu_i(\beta) \beta d\beta /\int^1_0
  \nu_i(\beta) d\beta.
  \label{5.12}
  \ee

  Our next step is the integration over $D\dot R$, which is done
  in the same way as in \cite{47}, and choosing the system where
  transverse total momentum  vanish, $\veP_\bot=0$, one obtains
   $$
   A_{lc} =\frac12 \int^T_0 dz_+
   \sum^3_{i=1} \left\{ \frac{m^2_i}{\mu_i}+ \int^1_0 d\beta
\frac{\sigma^2(r^{(i)}_\bot)^2
   }{\nu_i} +
    a_{3i} (\dot r_\bot^{(i)})^2-\right.
   $$
   \be
   \left.   -\lan \nu_i^{(2)}\ran
       \frac{(\dot r_\bot^{(i)}r_\bot^{(i)})^2}{(r_\bot^{(i)})^2}  -
     \frac{\lan \nu_i^{(0)}\ran a_{1i}
     (r^{(i)}_-+\lan \beta\ran_i \dot r^{(i)}_\bot r_\bot^{(i)})^2}
     {(r_\bot^{(i)})^2(a_{1i}-\lan \nu_i^{(0)}\ran)}\right\}.
      \label{5.13}
                \ee
            where we have defined
    \be
    \lan \nu_i^{(k)}\ran =\int^1_0 \nu_i(\beta) (\beta-\lan \beta
    \ran_i)^k d\beta, ~~
    \label{5.14}
    \ee

    Integration over $D\dot R_-$ with $\exp (iP_\bot \int^T_0 \dot
    R_- dz_+)$ (the exponent appearing in the standard way when going from
    Lagrangian to Hamiltonian representation) yields important
    constraint $\delta(a_1-P_+)$, i.e.
    \be
    a_1=P_+\label{5.15}
    \ee
    and integration over $DR_+$ is trivial, since $A_{lc}$ does not
    depend on $R_+$.

    Before doing calculations for the l.c. Hamiltonian, one should
    go over to the Minkowskian action, which is achieved by
    replacements
    $$
    \mu_i\to - i\mu_i^M, ~~\nu_i\to - i\nu_i^M,
    $$
    \be
    a_i\to - ia_i^M,~~ A\to -i A^M.
    \label{5.16}
    \ee
    Omitting the superscript $M$ in what follows one obtains for
    the Minkowskian action
    $$
    A^{(M)}_{lc} =\frac12 \int^T_0 dz_+ \sum^3_{i=1}
    \left\{-\frac{m^2_i}{\mu_i} +a_{3i} (\dot r_\bot^{(i)})^2-
    \int^1_0\frac{\sigma^2d\beta}{\nu_i} (r_\bot^{(i)})^2\right.
    $$
    \be
      \left.   -\lan \nu_i^{(2)}\ran
       \frac{(\dot r_\bot^{(i)}r_\bot^{(i)})^2}{(r_\bot^{(i)})^2}  -
     \frac{\lan \nu_i^{(0)}\ran a_{1i}
     (r^{(i)}_-+\lan \beta\ran_i (\dot r^{(i)}_\bot r_\bot^{(i)}))^2}
     {(r_\bot^{(i)})^2(a_{1i}-\lan \nu_i^{(0)}\ran}\right\}.
      \label{5.17}
      \ee

      From (\ref{5.17}) one can define in a direct way the l.c.
      Hamiltonian, writing
      \be
      A^{(M)}_{lc} =\int dt_+L^{(M)},~~ H=\sum\vep_\bot \dot{\veq}_\bot -L^{(M)}.
      \label{5.18}
      \ee
      One cannot choose $\veq_\bot^{(i)}$ strictly speaking as a
      set of canonical momenta for coordinates $\ver_\bot^{(i)}$,
      since the latter are not independent variables, subject
      according to (\ref{5.7}) to a condition
      \be
      \sum^3_{i=1} x_i\dot {\ver}_\bot^{(i)}=0.
      \label{5.19}
      \ee
      Instead the pair of coordinates $\dot{ \vexi}_\bot, \dot
      {\veta}_\bot$ is independent, and one can define canonical
      momenta $\vep^{(\xi)}_\bot, \vep^{(\eta)}_\bot$, as
      \be
      \vep^{(\xi)}_\bot=\frac{1}{i}
      \frac{\partial}{\partial\xi_\bot},~~ \vep_\bot^{(\eta)}
      =\frac{1}{i}\frac{\partial}{\partial\veta_\bot}.
      \label{5.20}
      \ee
      We can use nevertheless $\vep^{(i)}_\bot = \frac{1}{i}
      \frac{\partial}{\partial  {\ver^{(i)}}_\bot}$. Then
      \be
      \vep^{(\xi)}_\bot =\sum^3_{i=1} \vep^{(i)}_\bot c_{i\xi},~~
           \vep^{(\eta)}_\bot =\sum^3_{i=1} \vep^{(i)}_\bot c_{i\eta},
           \label{5.21}
           \ee
           where $c_{i\xi}, c_{i\eta}$ are listed from (\ref{5.7})
           $$
          c_{1\xi}=\left(\frac{x_3}{x_1+x_2}\right)^{1/2}=c_{2\xi},~~
      c_{3\xi}=-\left(\frac{x_1+x_2}{x_3}\right)^{1/2}
      $$
      \be
       c_{1\eta}=-\left(\frac{x_2}{x_1(x_1+x_2)}\right)^{1/2},~~
     c_{2\eta}=\left(\frac{x_1}{x_2(x_1+x_2)}\right)^{1/2},~~c_{3\eta}=0
\label{5.22}
     \ee

We are now in the position to use (\ref{5.18}) and calculate the
l.c. Hamiltonian, $$ H=\sum^3_{i=1} \{ \frac{m^2_i}{2\mu_i}
+\frac12\int^1_0 \frac{\sigma^2 d\beta}{\nu_i}
+\frac{(\vep_\bot^{(i)})^2-(\vep_\bot^{(i)}\ver_\bot^{(i)})^2/(\ver_\bot^{(i)})^2}{2a_{3i}}+
$$ $$ +\frac{\lan\nu_i^{(0)}\ran a_{1i}
(r_-^{(i)})^2}{2(r_\bot^{(i)})^2\mu_i}+
\frac{(p_\bot^{(i)}r_\bot^{(i)}+\frac{1}{\mu_i}\lan
\nu_i^{(0)}\ran  a_{1i} r_-^{(i)})^2\mu^2_i}{2(r_\bot^{(i)})^2
a_{3i} a^2_{2i}(2\mu_i-a_{2i})^2} \times$$
\be
[\mu_ia_{3i}+ \frac{(a_{2i}-\mu_i)^2}{\mu_i} (a_{3i}-2\mu_i)]\},
\label{89} \ee where $a_{ni}$ are defined  in (\ref{5.9}),
(\ref{5.9a}).

Momenta $\vep_\bot^{(i)}$ are not linearly independent, and from
(\ref{5.19}) expressing $\dot{\ver}_\bot^{(i)}$ through
$\vep_\bot^{(i)}$ one obtains the connection
\be
\sum^3_{i=1} \frac{x_i}{a_{3i}}(\vep^{(i)}-C\ver^{(i)}+
\ver^{(i)}D\frac{\vep^{(i)}\ver^{(i)}-C(\ver^{(i)})^2}{a_{3i}-(\ver_\bot^{(i)})^2
D}) = 0 \label{90} \ee where we have defined
\be
C=-\frac{\lan \nu_i^{(0)}\ran a_{1i}  r^{(i)}_-\lan
\beta\ran_i}{(\ver_\bot^{(i)})^2(a_{1i}-\lan \nu_i^{(0)}\ran)},
\ee
\be
D=\frac{\lan \nu_i^{(2)}\ran (a_{1i}-\lan \nu_i^{(0)}\ran) +\lan
\nu_i^{(0)}\ran a_{1i} \lan
\beta\ran^2_i}{(\ver_\bot^{(i)})^2(a_{1i}-\lan \nu_i^{(0)}\ran)}
\label{92} \ee

To understand better the structure of the Hamiltonian (\ref{89}),
consider first the limit of heavy quarks $m_i\gg\sqrt{\sigma}$, in
which case as was shown in \cite{47}, the inequality holds
$\mu_i\gg \nu_i, i=1,2,3.$ One has from (\ref{5.9}),(\ref{5.9a})
$a_{1i}= a_{2i}=a_{3i}=\mu_i$ and the Hamiltonian assumes the form
$$ H_{HQ} =\sum^3_{i=1} \left\{\frac{m^2_i}{2\mu_i}
+\frac12\int^1_0 \frac{\sigma^2 d\beta}{\nu_i}
+\frac{(\vep_\bot^{(i)})^2-(\vep_\bot^{(i)}\ver_\bot^{(i)})^2/(\ver_\bot^{(i)})^2}{2\mu_i}+\right.
$$
 \be \left. +\frac{(r_-^{(i)})^2}{2(r_\bot^{(i)})^2}\int^1_0
\nu_i d\beta +
 \frac{(p_\bot^{(i)}r_\bot^{(i)}+\lan\nu_i^{(0)}\ran
  r_-^{(i)})^2}{2(r_\bot^{(i)})^2\mu_i}
 \right\}.
 \label{93}
 \ee
 Introducing the dimensionless quantity
 $y_i\equiv\frac{\nu_i}{P_+}$ (which will be shown to be independent
 of $\beta$ and small, $y_i\ll 1$, one has from (\ref{5.10}) and
 (\ref{5.15})
 \be
 x_i=\frac{1}{1-Y} (\frac{\mu_i}{P_+}+\frac12 y_i), ~~
 \sum^3_{i=1}(\frac{\mu_i}{P_+}+y_i)=1
 \label{94}\ee
 where $Y=\frac12 \sum y_i$.

 This enables one to expand the mass term in (\ref{93}) around
 stationary points in $x_i$, and keeping the first order term in
 $y_i$ one obtains
 \be
 \sum^3_{i=1} \frac{m^2_i}{2\mu_i} =\frac{1}{2P_+}\{ M^2(1+2Y)
 +2M\sum(x_i-\frac{m_i}{M})^2\frac{1}{2m_i}\}\label{95}
 \ee
 where
 $M=\sum^3_{i=1} m_i$.

 We now define as in \cite{47} the $z$- component of momenta
 \be
 M(x_i-\frac{m_i}{M})\equiv p_z^{(i)}\label{96}
 \ee

Keeping now in expansion of (\ref{93}) only leading terms one
obtains $H_{HQ}=\frac{\mathcal{M}^2}{2P_+}$ with total mass
operator $$\mathcal{M}^2=M^2+2M \sum^3_{i=1}
\frac{(\vep^{(i)})^2}{2m_i} + M^2\sum^3_{i=1}
\frac{y_i}{(r_\bot^{(i)})^2} \left[(r_\bot^{(i)})^2+\right. $$
\be
\left.+\frac{(P_+r_-^{(i)})^2}{M^2} +
\frac{\sigma^2}{M^2}\left(\frac{(\ver_\bot^{(i)})^2}{y_i}\right)^2\right].
\label{97} \ee
 One can now define the stationary point of (\ref{97}) with
 respect to $y_i$,
 \be
 y_i^{(0)}=\frac{\sigma(\ver_\bot^{(i)})^2}{Mr^{(i)}},
 ~~(r^{(i)})^2=(\ver_\bot^{(i)})^2+ r_z^{(i)},~~ r^{(i)}_z\equiv
 \frac{P_+r_-^{(i)}}{M}.\label{98}
 \ee
Inserting $y_i=y_i^{(0)}$ back into (\ref{97}) one arrives at the
familiar nonrelativistic expansion
\be
\mathcal{M}^2 = M^2+2M\sum^3_{i=1} \left[
\frac{(\vep^{(i)})^2}{2m_i} + \sigma r^{(i)}\right].
 \label{99}
\ee Connection between $\vep^{(i)}$ also simplifies for $y_i\ll1$,
so that (\ref{90}) goes over into a simple relation $\sum^3_{i=1}
\vep^{(i)}=0$, as expected.

We now turn to the case of light quarks, where relations
(\ref{5.10}) and (\ref{5.15}) hold with $y_i$ nonzero, and observe
 that three strings contribute to the total momentum $P_+$ an amount,
$P_+^{str}\equiv \sum_i\int^1_0 \nu_i(\beta) d\beta$ which can be
significant and comparable to that of valence quarks,
$\sum^3_{i=1}\mu_i$. The numerical value of $\lan y \ran \approx
0.2$ obtained in \cite{48} for a light meson, suggests that a
larger value can be obtained for $P_+^{str}/P_+$, which can be
comparable to the 55\% of the energy-momentum sum rule, observed
in DIS experiment on nucleons. We suggest at this point following
\cite{48} that this 55
contribution, is mostly due to the string contribution $P_+^{str}$
from all Fock components of nucleon, most importantly from the
ground state strings, and from hybrid baryon excitation, where the
ratio $\frac{P_+^{str}}{P_+}$ is even larger.

This point will be elaborated elsewhere \cite{52}.

Another topic connected to the l.c. Hamiltonian (\ref{89}) is the
whole range of dynamical calculations similar to those done for
mesons in \cite{48}. One can solve for the eigenfunctions of
(\ref{89}) and calculate form-factor and valence part of structure
 function  of the  baryon-for ground and excited states. To
compute the full structure function however one needs higher Fock
components and first of all lowest hybrid excitations. Here comes
in the important problem of small $x$ - Regge-type behaviour and
connection of $t$-channel Regge poles  (including Pomeron) with
$s$-channel summation over baryon resonances (primarily hybrid
excitation) which is planned to discuss in \cite{52}.

\section{Discussion of the c.m. wave-function properties}

In this chapter we consider possible strategies and first
estimates in the determination of eigenvalues and eigenfunctions
 of the c.m. Hamiltonian (\ref{63}). The
latter is the sum of spin-independent part (the first four terms on
the r.h.s. of (\ref{63})) and $\Delta H_{spin}$, which is
calculated in \cite{40} and has a full relativistic Dirac
$4\otimes 4\otimes 4$ structure.

At this point one can apply two different approaches in treating
$\Delta H_{spin}$. In the most part of this section we shall
consider $\Delta H_{spin}$ as a correction which should be taken
into account in the first order of perturbation theory. This is
especially consistent for the perturbative part of $\Delta
H_{spin}$, which is known for light quarks only to the order
$O(\alpha_s)$. In the next section we also consider another
strategy, when $\Delta H_{spin}$, and especially its hyperfine
part, is treated in a full matrix form.

We start with the first approach and concentrate on the first term
$H_0$ in (\ref{63}), which is given in (\ref{60}). This term was
considered before in \cite{36} and numerical solution of the
$\Delta$-type states is presented there including the study of
Regge trajectories.

Since $H_0$ does not depend on spin and isospin and color degrees
of freedom are already integrated out,  one should look for fully
symmetric wave functions depending on spin variables $\sigma$,
isospin variables $\tau$, and coordinates $\vexi, \veta$.
Relativistic effects are taken into account in kinematics, where
einbein fields $\mu_i$ are introduced in (\ref{60}),and the latter
upon stationary point optimization in $\mu_i$, yield relativistic
energies as in (\ref{50}). However it is more advantageous to
solve (\ref{60}), which has nonrelativistic form without square
roots, and do optimization in $\mu_i$  for the resulting total
mass $M_0(\mu_i)$. The accuracy of this procedure for mesons was
checked in \cite{53} to be around or better than 5\%. This type of
procedure also simplifies calculation of all 4 corrections in
(\ref{63}), which contain $\mu_i$ explicitly.

Hence one can follow the construction of the fully symmetric wave
function as was done in  \cite{9,36}, which we slightly simplify
and adopt for notations used before. Namely Jacobi coordinates
$\vexi, \veta$ (\ref{41}) are chosen to be symmetric (s) and
antisymmetric (a) with respect to interchange indices 1 and 2, and
belong to the two-dimensional mixed representation of the
permutation group $S_3$, denoted $\psi^{\prime\prime}$ and
$\psi^{\prime}$  while one- dimensional ones are
$\psi^s$ and $\psi^a$. The same holds true for isospin wave
functions $\eta^{\prime\prime},\eta', \eta^s, \eta^a$ and
spin-isospin wave functions $\varphi^{\pp}, \varphi',
\varphi^s,\varphi^a$ and finally the full coordinate-spin-isospin
wave function which should be symmetric in interchange of all 3
indices is
\be
\Psi(\vez^{(i)},
\sigma,\tau)=\psi^s\varphi^s+\psi^a\varphi^a+\psi^{\pp}\varphi^{\pp}+\psi'\varphi'.
\label{100} \ee An additional requirement is that $\varphi^{(i)}$
and $\psi^{(i)},~~ i={\pp},\prime,s,a$, must belong to given total
angular momentum $L,m_L$ and total spin $S, m_S$ and isospin $I,
I_3$.

Inclusion of $\Delta H_{spin}$ helps to construct the wave
function as the eigenfunction of total momentum $J, m_J$.

Since the construction of spin-isospin functions for 3 quarks is
well-known \cite{1}-\cite{9}, we consicer here only the coordinate part
$\psi^{(i)} (\vexi, \veta)$. As in \cite{2}-\cite{4}, \cite{9,36}
we shall use the hyperspherical formalism \cite{54} which has
proved to be very accurate for the $3q$ ease, namely the lowest
hyperspherical function yields eigenvalues with 1

One can introduce hyperradius $\rho$ in the following way (note
 the difference from definition in \cite{36}, where the case of
equal masses $\mu_i$ was considered).
\be
\rho^2=\sum^3_{i=1} \frac{\mu_i(\vez^{(i)}-\veR)^2}{\mu} =
\vexi^2+\veta^2. \label{100a} \ee The coordinate wave function
$\psi(\vexi, \veta)$ can be expanded in an infinite series of
hyperspherical functions $u^\nu_K(\Omega)$ depending on  angular
variables $\Omega$, with $K$-- grand angular momentum, $K=L, L+2,
L+4,...$, and $\nu$ -- the set of all other quantum numbers, see
\cite{54} for a review,
\be
\psi(\vexi,\veta)=\frac{1}{\rho^2}
\sum_{K,\nu}u^\nu_K(\Omega)\psi^\nu_k(\rho).\label{101}\ee Writing
(\ref{60}) as
\be
H_0= \sum^3_{i=1}
(\frac{m^2_i}{2\mu_i}+\frac{\mu_i}{2})+h_0,\label{102} \ee one can
reduce the equation $h_0\psi=E\psi$ to a system of equations \beq
\frac{d^2\psi^\nu_K}{d\rho^2} +\frac{1}{\rho}
\frac{d\psi^\nu_K}{d\rho} +\left[ 2\mu E- \frac{(K+2)^2}{\rho^2}
\right] \psi^\nu_K=\\\nonumber
=2\mu\sum_{K',\nu'}U^{\nu\nu'}_{KK'}(\rho)
\psi^{\nu'}_{K'}(\rho),\label{103}\eeq where it was defined
\be
U^{\nu\nu'}_{KK'}(\rho)=\int u^{\nu +}_K(\Omega) V_{conf}
(\vexi,\veta) u^{\nu'}_{K'}(\Omega) d\Omega.\label{104}\ee The
confining potential $V_{conf}$ was considered in \cite{36}
 assuming linear $Y$-type form. The analysis
of matrix elements (\ref{104}) is also given in \cite{9,54}, and
here we shall use only the simplest form, namely the so-called
hypercentral component, which for the equal mass case
$(\mu_i=\mu)$ is
\be
U^{00}_{00}(\rho) = 1.118 \sqrt{2}\sigma \rho = 1.58 \sigma \rho.
\label{105}\ee The lowest order equation (\ref{103}) for $K=K'=0$
was solved numerically in \cite{9,2}. Below we shall demonstrate a
simpler approach which allows to obtain eigenvalues of this
equation analytically with accuracy of 1\% for lowest states. To
this end we take in  (\ref{103})  $K=K'$  (neglecting
nondiagonal coupling) and making a replacement $\psi^\nu_K(\rho)
=\frac{\bar\psi^\nu_K(\rho)}{\sqrt{\rho}}$, reduce equation to the form
\be
-\frac{1}{2\mu} \frac{d^2\bar \psi^\nu_K(\rho)}{d\rho^2} + W_{KK}
(\rho) \bar \psi^\nu_K(\rho) = E_{Kn} \bar
\psi^\nu_K(\rho)\label{106}\ee with
\be
W_{KK}(\rho) = b\rho+\frac{d}{2\mu\rho^2},~~
b=\sigma\sqrt{\frac{2}{3}}\frac{32}{5\pi},~~ d=\left
(K+\frac32\right) \left(K+\frac52\right).\label{107}\ee The
eigenvalue $E_{Kn}$ can be found using oscillator-well
approximation near the minimum of $W_{KK}(\rho)$
\be
\frac{dW_{KK}(\rho)}{d\rho}\left |_{\rho= \rho_0}=0,~~
\rho_0=\left(\frac{d}{\mu b}\right)^{1/3} \right.\label{108}\ee
which yields \be
E_{Kn} \cong W_{KK} (\rho_0)
+\omega\left(n+\frac12 \right)\equiv \frac{\sigma^{2/3}}{\mu^{1/3}} c_n
\label{109} \ee where for $K=0$
\be
W_{00} (\rho_0) =\frac32 \left( \frac{b^2d}{\mu}\right) ^{1/3}, ~~
\omega = \frac{\sqrt{3d}}{\mu\rho^2_0}.\label{110} \ee The
spectrum $\omega n$ in (\ref{109}) corresponds to the so-called "breathing
modes", when baryon is excited in its $\rho$ -dependent mode only.

Finally adding other terms in (\ref{60}) one has for $M_{Kn}$ - the
eigenvalue of $H_0$,
\be
M_{Kn} =\frac{3}{2} \mu +E_{Kn}(\mu)\label{111}\ee At this stage
one defines $\mu$ from the stationary point of (\ref{111}),
$\frac{\partial M_{Kn}}{\partial\mu}\left|_{\mu=\mu_0}=0\right.$,
which yields
\be
\mu_0=\left( \frac29 c_n\right)^{3/4} \sqrt{\sigma},~~ M_{Kn}
(\mu_0) =\sqrt{\sigma} 6\left(\frac29
c_n\right)^{3/4}.\label{112}\ee

The total spin-averaged mass of baryon corresponding to the
Hamiltonian (\ref{63}) is
\be
M^{tot}_{Kn}=M_{Kn} (\mu_0) +\lan \Delta H_{string} \ran +\lan
\Delta H_{coul}\ran +\lan \Delta H_{self}\ran.\label{119}\ee For
the lowest states with $L=0,1$ one can neglect $\lan \Delta
H_{string}\ran $, while the other two terms are
\be
\lan \Delta H_{self} \ran = -\frac{6\sigma}{\pi\mu_0}, ~~\lan \Delta
H_{coul} \ran =-\frac{\lambda
b^{1/3}}{(2\mu_0)^{2/3}\rho_0}\label{114} \ee where $\lan \Delta
H_{self}\ran $ is given in \cite{41}, while $(\Delta H)_{coul}$ is
in \cite{9,37} . Here notation is used
\be
\lambda = \alpha_s \frac83 \left( \frac{10\sqrt{3}
\mu^2_0}{\pi^2\sigma} \right)^{1/3}.\label{115}\ee Now $M_{Kn}
(\mu_0)$ is defined in (\ref{112}) and one should choose the only
input parameters (for light quarks we put all $m_i=0$) $\sigma$
and $\alpha_s$. The string tension $\sigma$ is renormalized due to
the presence of nondiagonal terms (\ref{23}) and therefore is
smaller than in the mesonic case (see \cite{46} for comparison
with lattice data and more discussion). For simple estimate below
we choose $\sigma =0.15$ GeV$^2$ (the same value as in \cite{8})
and take $\alpha_s=0.4$, which near its saturated value
\cite{55}.

Results of calculations  made according to Eqs.
(\ref{111})-(\ref{114}) are given in Table 1.\\

\begin{center}
{\bf Table 1}\\

Baryon masses (in GeV) averaged over hyperfine spin splitting for\\
$\sigma =0.15$ GeV$^2$,  $\alpha_s=0.4,~~ m_i=0$.\\

\begin{tabular}{|l|l|l|l|l|}
\hline State& $M_{Kn}+\lan \Delta H\ran_{self}$& $\lan \Delta
H\ran_{coul}$& $M^{tot}_{Kn}$& $M^{tot}(\exp)$\\\hline $K=0,n=0$&
1.36&-0.274& 1.08&1.08\\ $K=0,n=1$& 2.19&-0.274&1.91& ?\\
$K=0,n=2$&2.9&-0.274&2.62&?\\ $K=L=1,n=0$&1.85&-0.217&1.63&1.6\\
$K=2,n=0$&2.23&-0.186&2.04&?\\\hline
\end{tabular}
\end{center}

As it seen from the Table the calculated spin-averaged mass
$\frac12 (M_N+M_\Delta)=M^{tot}_{00}$ agrees well with the experimental
average, the same is also true for lowest negative party states
with $K=L+1$, which should be compared with $\frac{1^-}{2},
\frac{3^{-}}{2}$ states of $N$ and $\Delta$ respectively.

We also notice that breathing modes $(n>0)$ have excitation
energy around 0.8 GeV while orbital excitations
$K=L=1$ have energy interval around 0.5 GeV.

\section{Problem of spin-dependent forces}

We are now  turning to the spin-dependent interaction. For the
$3q$ case the corresponding nonperturbative and perturbative terms
are given in \cite{40}. They have been derived under the only
assumption of Gaussian dominance, i.e. only contribution  of the
bilocal correlator (represented by scalar functions $D$ and $D_1$)
was retained in (\ref{9}), Gaussian dominance being supported by
recent lattice data \cite{25,26}. The resulting spin-dependent
forces have in general the form of
 a product of two 4$\times$4 matrices,one for each interacting quark,
 and this is the most general  relativistic spin interaction.

 The expansion in powers of inverse quark mass was not used in
 \cite{40}, and for  light quarks the spin-dependent interaction
 is proportional to the  terms
  $\frac{1}{\mu_i\mu_j}$ and higher inverse mass terms,
  where $\mu_i$ are  are the same as in (\ref{102}) and
  (\ref{112}) and have  the meaning of constituent quark masses,
  which grow with excitation. For the lowest states $\mu_0\cong
  0.37$ GeV and grow fast with increasing $K,L$ and $n$.

  Now one could use two types of strategy to implement
  spin-dependent forces.

  1). Since all terms in (\ref{63}) except the last one $\Delta
  H_{spin}$  are diagonal in Dirac indices, then one  can
  calculate spin-independent wave-functions and account for spin
  effects calculating matrix elements $\lan \Delta
  H_{spin}\ran_{KLn}$. This procedure is actually used by most
  authors, and one can mention two positive moments associated
  with it. First of all in this procedure one treats
  spin-dependent forces as a perturbation, and it should work at
  least for high enough excitation, when $\mu_i\mu_j$ in the
  denominator become large. Secondly, the perturbative
  spin-dependent forces are known for light quarks only to the
  lowest order in $\alpha_s$ and therefore it is illegitimate to
  account those terms in higher than first order approximation.

  However doing so one immediately comes to a serious
  contradiction. Namely the theoretical estimates of perturbative
  hyperfine splitting for a reasonable value of $\alpha_s\approx
  0.4$ yield values around a hundred of MeV instead of 300 MeV for
  the $\Delta -N$ case \cite{9}. The phenomenological remedy  used
  is to  take $\alpha_s\sim 1$  and smearing the hyperfine
  $\delta$-function take the resulting potential to higher orders,
  which was criticized above.

  To resolve this contradiction it is suggested first of all to
  take into account the nonperturbative part of hyperfine
  interaction, which was derived in \cite{40}. It is known to
  yield a large part of  hyperfine splitting in light mesons
  \cite{27,28} and may be large also for baryons. Secondly, it is
  suggested in addition to use another strategy discussed below.

  2). In case when spin forces are important, as was discussed in
  the hyperfine case, one should take into  account that the same
  type matrix element which creates hyperfine splitting, also
  connects lower and higher components of Dirac bispinor.
  Physically this means excitation of negative energy components
  of quark wave function in baryon, which is also associated with
  the backward in time propagation of quarks.

  Therefore the strong hyperfine splitting implies also strong
  negative energy component excitation, and the solution of the
  total Hamiltonian (\ref{63}) should be sought  for in the form
  $$\Psi_B=\sum C^{\alpha\beta\gamma}_{ikl}
  \psi^{(i)}_\alpha\psi_\beta^{(k)}\psi^{(l)}_\gamma,
  $$ where $\alpha,\beta
\gamma$ are Dirac bispinor indices and $i,k,l$ refer to the
excitation state of a given quark.

This  strategy is equivalent to the full relativist 3-body
Bethe-Salpeter equation, which was studied in the quasipotential
form in \cite{56}.

Another possible treatment of the same problem was recently
initiated in \cite{38,39}, where 3-fold Dirac equations were
derived from the QCD Lagrangian for the baryon Green's function.

\section{Conclusions.}

We have derived the $3q$ Hamiltonian both in the c.m. and in the
l.c. coordinate systems. It was demonstrated that the c.m.
Hamiltonian can be written conviently as a sum of a main term
$H_0$ and four corrections in (\ref{63}), representing rotating
string energy, Coulomb energy, nonperturbative selfenergy
correction and spin interaction respectively. The explicit form of
all terms is given above, except for the last one, published
recently in \cite{40}.

The spin-averaged energy levels have been calculated analytically,
using hyperspherical formalism yielding accuracy around 1\% for
energy levels in linear confining potential. Results for $\Delta
-N$ system are in good agreement with experiment.The present paper
is meant to be a starting point of a new treatment of baryons,
where all types of forces are derived explicitly from the first
principles under the only assumption of the Gaussian dominance.
The spin-dependent forces derived for the first time in its
totality in \cite{40}  constitute an essential part of this new
approach.

\section{Acknowledgements}

The author has benefitted from a good working creative atmosphere
 in the Theory group of the Jefferson Lab.It is a pleasure to
thank all members of the Theory group,all staff and especially
Franz Gross for a kind hospitality. Many useful
discussions,remarks,suggestions are appreciated during talks and
contacts with V.Burkert, C.Carl\-son, R.Ed\-wards,J.Goi\-ty, F.Gross,
W.Mel\-nit\-chouk, I.Mu\-sa\-tov, A.Ra\-dyush\-kin, D.Ri\-chards,
W.Ro\-berts, C.Schat, R.Schia\-villa and J.Tjon. The partial
support of RFBR grants 00-02-17836,00-15-96736 and INTAS grants
00-110 and 00-366 is acknowledged.The author was supported by DOE
contract DE-AC05-84ER40150 under which SURA operates the TJNAF.\\

{\bf Appendix}\\

 \setcounter{equation}{0}
\renewcommand{\theequation}{A.\arabic{equation}}

 We start with the definition of the  string junction position
 $\vez^{(Y)}$ which is obtained from the minimum condition of the sum

 \be
 \sum^3_{i=1} | \ver^{(i)}| = \sum^3_{i=1} | \vez^{(i)}-\vez^{(Y)}|
 \label{A.1}
 \ee
and yields after differentiating in $\vez^{(Y)}$
 \be
 \sum^3_{i=1} \frac{\ver^{(i)}}{|\ver^{(i)}|} =\sum^3_{i=1} \ven^{(i)}=0.
 \label{A.2}
 \ee
 This implies that three unit vectors $\ven^{(i)}$ are at 120$^o$ with
 respect to each other, being in one plane. Therefore one can
 relate positions of quarks
 $\vez^{(ij)} \equiv \vez^{(i)} -\vez^{(j)}$ to  $\ver^{(i)}$ as follows
 \be (\vez^{(ij)})^2
=(\ver^{(i)})^2+(\ver^{(j)})^2+|\ver^{(i)}||\ver^{(j)}|, i\neq
 j=1,2,3.\label{A.3}
 \ee
 One can finally relate $\ver^{(i)}$to the Jacobi coordinates
 $\vexi,\veta$, Eq.(\ref{41})
 \be
 \ver^{(i)} = b_i\vexi + c_i\veta+\vedelta \label{a.3}
 \ee
 where coefficients $b_i,c_i$  are given in (\ref{41}) and
 $\vedelta=\veR-\vez^{(Y)}$ is to be found by solving (\ref{A.2}),
 (\ref{A.3}). Eqs. (\ref{A.3}) are algebraic and allow to find the
 lengths $r_i\equiv |\ver^{(i)}|, i=1,2,3$  through the quark
 positions $|\vez^{(ij)}|$, therefore $r_i$ will be assumed to be
 found explicitly. To find $\vedelta$ we place 3 quarks on the plane
 $x,y$ so that quarks 1 and 2 are on the $x$ axis. Assuming the
 $3q$ triangle to have all angles less than 120$^o$, one can
 compute both  components $\delta_x, \delta_y$ in terms of $r_i$.
 Thus the position of the string junction $\vez^{(Y)} (x_0, y_0)$ is
 found  to be
 \be
 y_0 =\frac{\sqrt{3} r_1r_2}{\sqrt{3r^2_2+(2r_1+r_2)^2}},
 x_0-x_1=\frac{(2r_1+r_2)r_1}{\sqrt{3r^2_2+(2r_1+r_2)^2}}\label{A.4}\ee
 where $(x_1,0)$ is the position of the quark 1.

 Similarly one obtains for $\delta_x,\delta_y$
\be
\delta_x=\frac{1}{3}\left[\left(\frac{r_2+r_3}{2}-r_1\right)
(2r_1+r_2) +3(r_2-r_3)
r_2\right]\frac{1}{\sqrt{3r^2_2+(2r_1+r_2)^2}}\label{A.5} \ee
\be
\delta_y =\frac{r_3(r_1+r_2)-2r_1r_2}{\sqrt{3}
\sqrt{3r^2_2+(2r_1+r_2)^2}}.\label{A.6}\ee One can see that for
the symmetric case $r_1=r_2=r_3$ the  string junction and the c.m.
positions coincide, $\delta_x=\delta_y=0$.

Finally we quote for the convenience of the reader the  expression
of the sum (A.1) through the quark positions only, taken from
\cite{36}.
\be
\sum^3_{i=1} r_i =\frac{\sqrt{3}}{2}[(\vez^{(12)})^2 +3
(\vez^{(3)}-\veR)^2+ 2\sqrt{3}|\vez^{(12)}\times
(\vez^{(3)}-\veR)|]^{1/2}.\label{A.7} \ee

\end{document}